# Energy Limits to the Gross Domestic Product on Earth

Andreas M. Hein[1,2], Jean-Baptiste Rudelle[3]

[1] Université Paris-Saclay, CentraleSupélec, Laboratoire Genie Industriel, 3 rue Joliot-Curie 91190 Gif-sur-Yvette, France, andreas-makoto.hein@centralesupelec.fr

[2] Initiative for Interstellar Studies, 27-29 South Lambeth Road, South Lambeth Road, London SW8 1SZ, United Kingdom

[3] Zenon Research Center, 16 rue Seguier, 75006 Paris, France

## Abstract

Once carbon emission neutrality and other sustainability goals have been achieved, a widespread assumption is that economic growth at current rates can be sustained beyond the 21st century. However, even if we achieve these goals, this article shows that the overall size of Earth's global economy is facing an upper limit purely due to energy and thermodynamic factors. For that, we break down global warming into two components: the greenhouse gas effect and heat dissipation from energy consumption related to economic activities. For the temperature increase due to greenhouse gas emissions, we take 2 °C and 5 °C as our lower and upper bounds. For the warming effect of heat dissipation related to energy consumption, we use a simplified model for global warming and an extrapolation of the historical correlation between global gross domestic product (GDP) and primary energy production. Combining the two effects, we set the acceptable global warming temperature limit to 7 °C above pre-industrial levels. We develop four scenarios, based on the viability of large-scale deployment of carbon-neutral energy sources. Our results indicate that for a 2% annual GDP growth, the upper limit will be reached at best within a few centuries, even in favorable scenarios where new energy sources such as fusion power are deployed on a massive scale. We conclude that unless GDP can be largely decoupled from energy consumption, thermodynamics will put a hard cap on the size of Earth's economy. Further economic growth would necessarily require expanding economic activities into space.

## 1 Introduction

Economic growth, expressed via Gross Domestic Product (GDP) growth, is at the center of economic policy, as it is widely considered as intrinsically desirable (Brinkman and Brinkman, 2011; Lewis, 2013). GDP is positively correlated with life expectancy, safety, peace, and what is generally considered a higher quality of life for the majority of people (Lewis, 2013). Although its critics have contested that GDP is an appropriate indicator for social welfare (Costanza et al., 2014; Kallis, 2011; Kubiszewski et al., 2013; Van den Bergh and Kallis, 2012), there is a broad consensus that it should or can be sustained long-term. For example, the latest IPCC reports are based on the Shared Socioeconomic Pathways (SSPs) scenarios, which all assume economic growth at least up to the year 2100 (Leimbach et al., 2017; O'Neill et al., 2017; Van Vuuren et al., 2011). This consensus seems to be further confirmed by the limited traction of a debate on negative growth scenarios (Hickel and Kallis, 2019; Kuhnhenn, 2018).

While economic growth is considered desirable, it is correlated with negative environmental impacts such as greenhouse gas emissions (Csereklyei et al., 2016; Heil and Selden, 2001; Sorrell, 2015; Stern, 2018). Hence, a major concern of policy-making is how economic growth and environmental protection can be reconciled. The reconciliation of both is commonly termed "green growth" and is being promoted by major international organizations such as the United Nations and the World Bank (Ekins, 2000; Hallegatte et al., 2011; Jänicke, 2012). However, it is unclear whether green growth is feasible on a long term time scale.

The first obstacle to green growth is that on a global scale, avoiding environmental impact from economic activities seems to be challenging, and the empirical underpinnings for its feasibility in practice are rather weak (Hickel and Kallis, 2019; Ward et al., 2016). Nevertheless, we assume that green growth is feasible within planetary boundaries, including the ability to achieve carbon neutrality and to decouple economic growth from commodity extraction (Rockström et al., 2009; Steffen et al., 2015).

However, we argue that even green growth cannot be sustained indefinitely due to a fundamental physical limit, which is energy (Ayres, 1996; Douglass et al., 2004; Knox, 1999; Mayumi et al., 1998; Pearce, 2008). The first limit to energy could be imposed by its supply. The energy supply limit appears in different forms related to each specific energy source.

A second limit to energy could be imposed by its consumption, resulting in heat dissipation. Today, heat dissipation from energy consumption is a very small contributor to global warming compared to greenhouse gas emissions (Knox, 1999; Pearce, 2008). However, if energy consumption keeps growing, this contribution could become much more significant. A common denominator of these publications is that the related limits are considered to be so far in the future that they have no practical implications. So far, it seems that no publication has estimated limits to the size of the economy related to energy constraints and estimated when these limits might be reached. This article argues that these limits will be reached soon enough to have potentially significant long-term policy implications. Specifically, it might imply that the current focus on green growth as a solution to sustaining growth might not be sufficient long-term, and further measures need to be taken.

In this paper, we aim at establishing long-term limits of the global GDP, based on energy supply and heat dissipation considerations. The objective is to obtain rough, order of magnitude estimates. For this purpose, we use a simplified model for global warming and an extrapolation of the historical correlation between global gross domestic product (GDP) and primary energy production. We further develop four scenarios based on the viability of large-scale deployment of carbon-neutral energy sources and estimates for the temperature increase due to greenhouse gas emissions.

## 2 Methodology

For estimating energy-related limits, we first develop a relationship between GDP and energy, which we will extrapolate into the future. Subsequently, we introduce different energy sources, namely, renewables and non-renewable energy sources. To determine the heat dissipation limit, we use a simplified model for global warming and estimate an upper limit to acceptable global warming. We further develop four energy scenarios and estimates for the temperature increase due to greenhouse gas emissions.

### 2.1 Economic growth and energy

Theoretical underpinnings for the relationship between energy consumption and economic activities have been developed in the ecological economics literature (Ayres, 1997, 1999, 1996; Georgescu-Roegen, 1993, 1975; Stern, 2018). The empirical correlation between GDP growth and energy consumption has been studied extensively, although there is an ongoing discussion on the direction of the causal relationship (Bruns et al., 2014; Costantini and Martini, 2010; Payne, 2010). The most extensive meta-analysis to date concludes that GDP growth causes energy consumption while controlling for energy prices (Bruns et al., 2014). In the following, we refer to the theories from the ecological economics literature as well as the empirical results such as Bruns et al. (2014) and assume that every economic activity requires a certain level of energy consumption (Stern, 1997). It follows that economies with a higher GDP also tend to consume more energy. In other words, if there is a specific upper limit to energy consumption on Earth, it would also create an upper limit for GDP.

Due to losses in production and transportation, energy consumption is always smaller than energy production and for our purpose, we will focus in the following on energy production. To determine the relationship between energy production and GDP, we use historical data from 1960 to 2015 (World Bank, 2019) and global primary energy production data (Ritchie and Roser, 2019), normalized by the first year in the data set (1960). We use the MATLAB curve fitting tool and a power law fit ($R^2 = 0.99$), resulting in:

$$\frac{G(t)}{G(t_0)} = 0.915 \left[\frac{E(t)}{E(t_0)}\right]^{1.48} \tag{1}$$

where $G$ and $E$ represent respectively the world GDP and annual energy production for year t and $t_0 = 1960$. Figure 1 shows the data points and the power law fit, which seems to capture the overall trend.

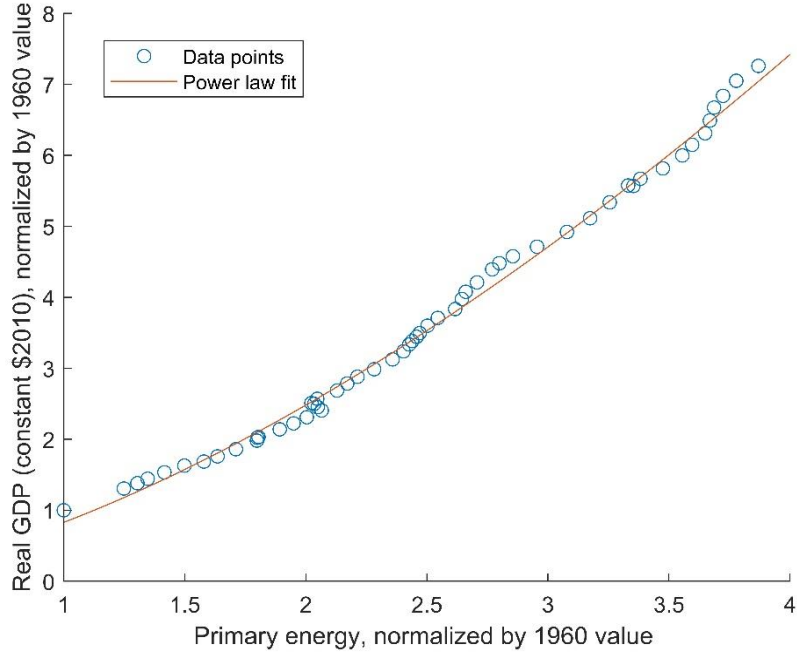

**Figure 1: Data points for normalized world primary energy generation versus constant $2010 GDP from 1960 to 2015 and power law fit**

In the following, we assume that this relationship holds several centuries into the future. For further calculations, we take $t_0 = 2017$ and $E(t_0) = 162,000$ TWh. The multiplicative correlation coefficient needs to be adjusted accordingly and we obtain 0.928 for the new reference year.

## 2.2 Limits to renewable energies

To distinguish between energy sources which contribute to global warming via heat dissipation and those who essentially transform solar energy without long-term storage, we are separating energy production into two components:

$$\frac{G(t)}{G(t_0)} = 0.928 \left[\frac{E_r(t) + E_h(t)}{E(t_0)}\right]^{1.48} \tag{2}$$

$E_r$ denotes the energy supplied directly or indirectly by solar radiation during a given year. This includes solar panels, wind turbines, biomass and hydroelectricity. These energy sources have in common that they use the incoming solar radiation during a given year. This means that whatever heat dissipation they generate, it is already accounted for in the existing thermodynamic equilibrium of the Earth biosphere and no additional heat dissipation is generated.

$E_h$ represents all other energy sources, mostly fossil fuels and nuclear with a very small contribution of geothermal. All these energy sources have in common that heat dissipation related to their usage is additive to the existing thermodynamic equilibrium of the Earth's biosphere.

To determine the upper limit of global GDP, we therefore have to determine upper limits for $E_r$ and $E_h$. $E_r$ is fully renewable energy, so there is no risk of depletion. The solar flux density incident on a plane surface at Earth's distance from the Sun is on average 1365 Watts per square meter (Kopp and Lean, 2011) and about 341 W/m² when averaged-out over the Earth's surface (Knox, 1999). The total incident power on the surface area of a disk with the diameter of Earth is 174 PW. On an annual basis, the equivalent incident solar radiation energy is $R_{sol} = 1.5$ Billion TWh. This would suggest that the theoretical upper limit is $E_{r\,lim} = 0.7\,R_{sol}$, taking the Earth's atmosphere into account, which reflects about 30% of the incoming solar radiation (Szargut, 2003).

Our ability to transform solar radiation into usable energy is directly proportional to the surface area we dedicate to it. As a result, the actual limit for $E_r$ is less about raw potential or even technological development, but more about practical constraints, above all competition from other use of real estate. Numerous studies have been conducted to estimate the limits of solar, wind, and hydropower, some with upper limits in the thousands to tens of thousands of EJ/a (De Vries et al., 2007; Defaix, 2009; Deng et al., 2015; Edenhofer et al., 2011; Goldemberg, 2000; Hoogwijk, 2004; Jacobson and Delucchi, 2011; Krewitt et al., 2009; Siegfriedsen et al., 2003; Trieb et al., 2009). More recently, Deng et al. (2015) provided a more precise estimate, based on surface area use on the km-scale. They determine a limit for renewable energy sources equivalent to about 200,000 to 1 million TWh, which includes solar, wind, and hydropower plus a smaller contribution from biomass. This is 0.02% to 0.1% of the theoretical limit mentioned in the previous paragraph. Ultimately, the difference between the two estimates will depend on how far our global civilization will favor renewable energy production versus competing usages of the real estate. In the result section, we will show the difference of impact if we take $E_{r\,lim} = 200,000$ TWh (lower bound) or $E_{r\,lim} = 1$ million TWh (upper bound).

## 2.3 Other energy sources

In the following, we assume that energy from fossil fuels will be largely substituted within the coming centuries, primarily due to measures taken to reduce greenhouse gas emissions. As a result, we consider their long-term contribution to $E_h$ as negligible.

Beyond fossil fuels, the main other option available to produce energy at scale is nuclear fission reactors. Today the lack of broad public support for this technology seriously limits its expansion. However, given the increasing pressure to reduce fossil fuels in our energy mix, we might see a reacceleration of new nuclear plant construction programs in the future.

For long time horizons, we have to consider that other new types of energy sources could also become available in the future. There are two technologies in particular that have the potential, at least in theory, to provide massive additional energy supply:

1) Nuclear fusion reactors:
   At this stage, different fuels are considered for fusion reactors, including deuterium that can be easily extracted from seawater. Although the long term requirements for fusion are not fully understood yet (Bradshaw et al., 2011), assuming economic viability for this technology, it could be considered as an "unlimited" energy source for the future.

2) Solar power satellites:
   If we could harvest solar energy directly in space and beam it back to be consumed on Earth (Bergsrud et al., 2013; Glaser et al., 1998; Hendriks et al., 2004), it would give access to practically limitless real estate to install solar panels. Assuming that we will be able to scale and operate such a technology in practice, there is no obvious limit to the amount of energy that could be beamed back to power Earth's economy (Hendriks et al., 2004).

Assessing the feasibility of fusion reactors or solar power satellites is well beyond the scope of this study. However, we argue in the following section that even if we benefit from such new "unlimited" energy sources, we will still face limits to the usage of their energy on Earth due to thermodynamics.

## 2.4 Thermodynamic limits to energy production

As discussed in Section 2.1, the consumption of energy to perform economic activities results in heat dissipation. Several models for estimating global temperature increase from heat dissipation have been proposed in the literature (Knox, 1999; Rose, 1979). We use the linear global warming model proposed by Knox (1999), which correlates surface heat release with global warming via a two-layer earth model. This model does not take into account global warming effects from changes in the atmospheric composition, such as from greenhouse gas emissions. It also assumes that surface heat release does not influence optical parameters such as surface and atmospheric reflectivity. We use this model due to its simplicity and its ability to still capture key parameters of global warming. $\Delta T_h$, the average temperature increase at the Earth's surface due to heat dissipation from human activities is expressed as:

$$\Delta T_h = C \, \varepsilon(t) \tag{3}$$

where $C$ is the temperature constant 114 K and $\varepsilon(t)$ the rate of surface heat release from human activities at the time $t$ in units of the average incident solar radiation $R_{sol}$. To assume that $C$ is constant, we make the hypothesis that we can neglect the change of atmospheric and surface optical parameters due to the additional surface heat release. This makes this equation valid for small variations of $\varepsilon(t)$. Given that anthropic heat dissipation will remain a relatively small fraction of the total incident solar radiation, we consider this condition satisfied. To evaluate the sensitivity of our assumption, we will show in the result section the impact of less optimistic values of $C$. We assume that heat release from human activities is equal to energy production, as it is either lost due to waste heat or transformed into useful work. Translated into thermodynamic terms, this means that if a process is irreversible, exergy (amount of useful work) destruction results in heat dissipation. In practice, all economic processes are considered irreversible and therefore any form of extracting useful work from an energy source leads to heat dissipation (Georgescu-Roegen, 1993).

As we have seen, all heat dissipation related to $E_r$ is already included in the existing thermodynamic equilibrium of the Earth biosphere. So we need to apply the increase of temperature due to heat dissipation to $E_h$ only. We also assume that heat dissipation from energy production related to economic activities can be averaged over the Earth's surface area (due to atmospheric re-equilibrium). Hence, $\varepsilon$ can be expressed as:

$$\varepsilon(t) = \frac{E_h(t)}{R_{sol}} \tag{4}$$

In 2017, $\varepsilon$ was roughly equals to $\frac{E(t_0)}{R_{sol}} = 0.011\%$, based on energy data from Ritchie and Roser (2019) and the solar constant of 1365 W/m² from Kopp and Lean (2011), multiplied by the Earth's surface. This means that global warming from heat dissipation accounts for $\Delta T_h = 0.012$ K. This is two orders of magnitude lower than the warming effect from anthropic greenhouse gas emissions.

From equations (2), (3), and (4), we can now write the GDP ratio as:

$$\frac{G(t)}{G(t_0)} = 0.928 \left[ \frac{E_r(t)}{E(t_0)} + \frac{\Delta T_h}{C} \left( \frac{R_{sol}}{E(t_0)} \right) \right]^{1.48} \tag{5}$$

We define $\Delta T$ as the global temperature increase resulting from two effects:

- The heat dissipation $\Delta T_h$ described above,
- The temperature increase $\Delta T_C$ due to greenhouse gas emissions.

We denote $\Delta T_{lim}$ occurring at the time $t_{lim}$, the maximum permissible $\Delta T$ temperature increase on Earth. Beyond that limit, the damage would result in a collapse of the world economy.

We define $\Delta T_{Cmax}$ as the maximum temperature increase due to the greenhouse effect of anthropic greenhouse gas emissions. This temperature increase is considered to be irreversible for at least 1,000 years after emissions stop (Archer and Brovkin, 2008; Eby et al., 2009; Solomon et al., 2009). In principle, negative emission technologies could remove atmospheric greenhouse gasses and thereby reduce the temperature (Cao and Caldeira, 2010; Tokarska and Zickfeld, 2015). A reduction of the average temperature by a fraction of a Kelvin would require negative emission technologies to be deployed on a massive scale for at least a century (Fuss et al., 2014; Tokarska and Zickfeld, 2015). However, more recent models representing the feedback loops between the atmosphere and other Earth system elements such as the oceans suggest that the effect of negative emission technologies on average temperature might be very limited over the timeframes considered in this study (Rickels et al., 2018). In the presence of these uncertainties, we assume that $\Delta T_{Cmax}$ is a constant value.

We also assume that all energy production in the future will be carbon neutral. As a result, the effect of greenhouse gas and heat dissipation are independent and additive. We can therefore write $\Delta T_{lim}$ as:

$$\Delta T_{lim} = \Delta T_h + \Delta T_{Cmax} \tag{6}$$

From Equations (5) and (6), we can therefore derive the upper limit of the GDP ratio:

$$G_{lim} = \frac{G(t_{lim})}{G(t_0)} = 0.928 \left[ \frac{E_{r\,lim}}{E(t_0)} + \frac{(\Delta T_{lim} - \Delta T_{Cmax})}{C} \left( \frac{R_{sol}}{E(t_0)} \right) \right]^{1.48} \tag{7}$$

## 2.5 Estimating $\Delta T_{lim}$ the maximum permissible temperature increase on Earth

First, we must admit that as of today, the degree of global warming resulting in an actual existential threat to humanity and life on Earth is not well understood.

The most extreme scenario called *runaway greenhouse* is when the entire ocean evaporates. A slightly less extreme scenario termed *moist greenhouse*, is where the planet would gradually lose its water to space, which would be unacceptable. However for this to happen, it is believed that the Earth surface temperature might have to exceed 340 K (Ramirez et al., 2014). The consensus is that for the temperatures we are considering in this study, a *runaway or moist greenhouse* is very unlikely (Goldblatt and Watson, 2012).

However, we do not need a full *runaway or moist greenhouse* to make global warming unacceptable. It has been argued that there is an upper limit to how much the human body can adapt to heat, quantified by a maximum acceptable wet-bulb temperature (Sherwood and Huber, 2010). Any exceedance of 35 °C for extended periods should induce dangerous hyperthermia in humans and other mammals. Sherwood and Huber (2010) estimate that this would start to occur in some part of the world with an average global increase of 7 K and would make the impacted regions hardly habitable. Other authors estimated that a global temperature increase beyond 5 K may lead to an existential threat for human civilization, although the impact is not well explored (Xu and Ramanathan, 2017). Even if they are hard to quantify, commonly cited threats include rising sea level and more frequent and bigger natural disasters that could jeopardize further economic growth (IPCC, 2014). It has also been estimated that about 90% of the existing species may die out at a 6 K increase (Barnosky, 2014). The economic impact of global warming, in particular for extreme temperature increases, is not well understood and seems to be grossly underestimated in existing models (Wagner and Weitzman, 2015; Weitzman, 2011). Weitzman (2011) refers to Sherwood and Huber (2010) to make the point that such temperature levels would be an extreme threat to human civilization but admits that quantifying the damage would be difficult. The rate at which temperature increases is also an important risk factor. A rapid warming over the course of a few decades only, would put even more stress on our ecosystem and our economy (Barnett et al., 2015; Klein et al., 2014).

Furthermore, we assume via Equation (3) a linear relationship between global temperature increase and surface energy release. We, therefore, exclude tipping points up to $\Delta T_{lim}$, which would abruptly increase the temperature, such as the release of massive amounts of methane from permafrost (Koven et al., 2011; MacDougall et al., 2012).

For our purpose, we just need to estimate a point where the heat stress would be so great that it would prevent any further economic growth. This is necessarily a much lower level than what would cause an actual existential risk for humanity. From the survey of GDP impact of climate change in Tol (2018), we conservatively estimate that at $\Delta T_{lim} = 7$ K (equivalent to an average surface temperature of 295 K), global warming would cause too much damage to allow the world's economy to grow any further. This being said, as some future unknown technology might be capable to mitigate the effects of global warming, we run some sensitivity analysis to see what additional GDP growth we would get if mankind was willing to experiment going up to $\Delta T_{lim} = 10$ K.

## 2.6   Four Scenarios for the future of Earth's economy
To illustrate potential paths for the Earth's economy, we have synthetized the results into four polarized scenarios. This classification does not claim to be exhaustive and other inputs in equation (7) could create many other valid paths. However, our objective is to explore extreme cases and thereby take into account the uncertainties in the input variables. We argue that a scenario-based analysis is more appropriate in this case than a traditional sensitivity analysis or probabilistic methods, as our aim is rather to explore possible futures, including extreme cases, than finding the most likely outcome.

Our scenarios are positioned with respect to two underlying dimensions:

1) *Climate change mitigation and energy transition measures:*
   How effectively climate change mitigation measures are implemented will determine the maximum temperature increase due to greenhouse gas emissions after the point where carbon neutrality is reached. More effective measures will lead to a lower temperature increase, while less effective measures will lead to a higher temperature increase. In parallel, the energy supply limit depends on the ability to exploit the full potential of renewable energy sources.

2) *Developing and scaling up new carbon-neutral energy sources:*
   In contrast to current renewable energy sources, these new types of energy sources are assumed to be carbon neutral and without practical fuel supply constraints. Candidate technologies are nuclear fusion and solar power satellites. The basic principles of these technologies are well understood, however, their commercialization at large-scale remains uncertain (Cardozo, 2019; Gi et al., 2020; Hendriks et al., 2004).

For each of these two dimensions, we defined two cases. Regarding climate change mitigation and energy transition measures, we distinguish between the case where these measures are implemented poorly and the case where they are implemented strongly. For this purpose, we take the stabilized global temperature increase from pre-industrial levels from different scenarios in the literature (Collins et al., 2013; Solomon et al., 2009). In the following we only focus on those scenarios where climate mitigation measures have been successfully put in place within the 21st century. For most scenarios, stable temperatures are achieved between the year 2100 and 2150. For the poor implementation, we assume that global warming due to greenhouse gas emissions leads to an average temperature increase of 5 K compared to pre-industrial levels. This value is consistent with more pessimistic scenarios in the literature (Collins et al., 2013; Solomon et al., 2009).  For a strong implementation, we essentially assume that the 2 K goal, which is currently considered to be achievable, has indeed been achieved (Rogelj et al., 2016). The bounds on renewable energy $E_{r\,lim}$ are defined by the lower and upper bounds given in Deng et al. (2015). We treat the bounds on renewable energy and the average temperature increase as independent variables. As a result:

for poor measures, we use $E_{r\,lim} = 200.000$ TWh and $\Delta T_{Cmax} = 5$ K

for strong measures, we use $E_{r\,lim} = 1$ million TWh and $\Delta T_{Cmax} = 2$ K

Regarding new carbon-neutral energy sources, we assume for the first case that new carbon-neutral energy sources are not available on a large scale, which might occur, for example, if fusion power and solar power satellites cannot be scaled up to provide significant fractions of the primary energy supply. For this case, the bounds on available primary energy are given by the bounds on renewable energy sources ($E_{h\,lim} = 0$). In the second case, we assume that new carbon-neutral energy sources are available and can be scaled up to the global warming limit.

The four scenarios are defined, based on a combination of the two cases for each dimension.

# 3 Results

## 3.1 Maximum size of Earth's economy

Based on Equation (7) and the four scenarios presented in section 2.6, we obtain the results in Table 1 for the maximum global GDP divided by the 2017 global GDP, denoted as $G_{lim}$.

**Table 1 : GDP limit with respect to 2017 global GDP for the four scenarios, assuming a maximum acceptable temperature increase of $\Delta T_{lim}$ of 7 °C.**

| Carbon-neutral energy breakthrough<br><br>Climate change mitigation and energy transition | No | Yes |
|---|---|---|
| Poor | **Scenario 1**<br>$G_{lim} = 1.3$ | **Scenario 3**<br>$G_{lim} = 1784$ |
| Strong | **Scenario 2**<br>$G_{lim} = 14$ | **Scenario 4**<br>$G_{lim} = 7003$ |

The results indicate that there is a difference of three orders of magnitude for $G_{lim}$ between the case where new carbon-neutral energy sources are available and can be scaled-up or not. The effect of climate change mitigation and energy transition measures is comparatively small but is still about one order of magnitude.

A dark variation of Scenario 3 would be if mankind was willing to push $\Delta T_{lim}$ aggressively from 7 K to 10 K. This extra temperature buffer would allow in theory to get the GDP to the same level as what can be achieved through Scenario 4. However, the drastic reduction of planetary habitability would increase dramatically the risk of an economic collapse way before the limit is reached (Sherwood and Huber, 2010).

It is also interesting to note from Equation (7) that if $\Delta T_{lim} - \Delta T_{Cmax}$ converges towards zero, scenarios 3 and 4 collapse respectively into scenarios 1 and 2. For instance, if after further studies, we realize that $\Delta T_{lim}$ has to be set at 5 K maximum, then it means that in scenario 3, we get $E_{h\,lim} = 0$. In other words, too much global warming from greenhouse gas emissions would put humanity into a situation, where the Earth's economy will not be able to benefit from energy breakthroughs such as fusion power and will be capped to what existing renewable energy sources can provide.

As discussed earlier, Equation (3) makes several simplifications, which lead to a low value for the constant *C*. Knox (1999) proposed alternative global warming models which imply higher values for the temperature constant *C*. If we plug in those variations, the GDP upper limits are significantly lower for scenarios 3 and 4, as shown in Table 2.

**Table 2 : $G_{lim}$ values for alternative global warming models in Knox (1999)**

|  | Model Constant C [K] | **Standard** | Calculation1 | Calculation2 | Calculation3 |
|---|---|---|---|---|---|
|  |  | **114** | 122 | 126 | 154 |
| Scenario 3 | $G_{lim}$ | **1 784** | 1 615 | 1 540 | 1 148 |
| Scenario 4 | $G_{lim}$ | **7 003** | 6 344 | 6 053 | 4 523 |

## 3.2   Duration to upper GDP limit

The time it takes to reach the GDP upper limit depends on its future annual growth rate. GDP extrapolations beyond the year 2100 seem to be hard to find. Long-term GDP growth rates up to the year 2100 are estimated to be around 2% per year with an uncertainly of +/- 1% (Christensen, 2017). We courageously extrapolate these values centuries into the future. We use 2017 as the base year. The results for 1, 2, and 3% annual GDP growth are shown in Table 3, where we call these growth rates respectively low, medium, and high.

**Table 3: Years into the future (from 2017) and actual year for which the GDP limits will be reached.**

|  |  | **GDP annual growth rate** | | |
|---|---|---|---|---|
| **Years to** | $G_{lim}$ | High 3.0% | **Medium 2.0%** | Low 1.0% |
| Scenario 1 | 1.3 | 8 | 13 | 25 |
| Scenario 2 | 14 | 90 | 134 | 267 |
| Scenario 3 | 1 784 | 253 | 378 | 752 |
| Scenario 4 | 7 003 | 300 | 447 | 890 |
| **Date to** | $G_{lim}$ |  |  |  |
| Scenario 1 | 1.3 | 2 025 | 2 030 | 2 042 |
| Scenario 2 | 14 | 2 107 | 2 151 | 2 284 |
| Scenario 3 | 1 784 | 2 270 | 2 395 | 2 769 |
| Scenario 4 | 7 003 | 2 317 | 2 464 | 2 907 |

For scenario 1, it is quite certain that the GDP will grow beyond 30% in the near future due to the continued use of fossil fuels. This implies that to reach carbon neutrality and exit the fossil fuel economy might lead to a decrease in GDP, in particular if carbon capture and storage technologies are not deployed on a massive scale (Hickel and Kallis, 2019). Without a technology breakthrough in carbon-neutral energy sources (Scenarios 1 and 2), the upper limit of GDP will be reached within the 22[nd] century for medium and high growth. For low growth, the limit will be reached in the 23[rd] century.

It is interesting to note that even with a technology breakthrough that would provide us with practically unbounded carbon-neutral energy, the upper limit of Earth's GDP would be reached in about four centuries for medium growth. If we consider that hopefully our civilization will last for multiple millennia, this is a fairly short window for GDP growth. As noted in the previous section, if greenhouse gas warming brings us by itself too close to the temperature maximum, we could be in a situation where we could not take advantage of any new energy source, regardless of our ability to mature and scale it.

# 4 Discussion

An important implication of our results is that even if carbon neutrality is reached and humanity succeeds in living within the planetary boundaries (Rockström et al., 2009), economic growth at current rates will come to an end in a time frame shorter than a few centuries. At least on this timescale, this result is in contradiction with the idea that economic growth can be sustained, if only natural resources are used sustainably (Ekins, 2000; Hallegatte et al., 2011; Jänicke, 2012). These limits will be reached within the lifetime of a few generations into the future, even for medium growth. These generations will have to face both, the consequences of global warming and a stagnant, energy-constrained economy. In the following, we will discuss limitations to the validity of our results.

There are several sources of uncertainty in our model, such as the rate of GDP growth, correlation between heat dissipation and climate change, acceptable global warming limit, and extrapolation of the energy-GDP relationship. Due to these large uncertainties, we opted for a scenario-based approach to identify extreme cases, rather than finding the most probable outcome. This explains that we have a difference of several orders of magnitude in the size of the economy and the year in which the limit would be reached. Our point is that even if we look at these extreme cases and the fact that the size of the economy differs by several orders of magnitude, the timeframes in which economic growth can happen are still limited to a few centuries. It is also clear that our results would not be valid if decoupling between energy consumption and GDP can be achieved on a global scale (Ward et al., 2016). We will address the decoupling argument in more detail in the following section 4.1.

We have previously selected a 7 K temperature increase as our upper limit, as it implies an existential threat to the habitability of the planet for humans. Due to the damage incurred by global warming, economic growth could slow down and stagnate way before that limit (Barnosky, 2014; Xu and Ramanathan, 2017). This again implies that our 7 K bound is rather a conservative estimate.

While we have decided to focus on energy constraints, there are of course many other factors that could limit global GDP, such as diminishing returns on technology improvements, over-pollution or shortage of critical commodities. All these factors would contribute to a lower GDP upper limit than presented here, which in return means that the upper limit we presented here is an optimistic estimate.

On the normative side, our basic assumption in this paper is that economic growth is somewhat desirable. Obviously, if de-growth is an acceptable scenario, we do not have to worry much about GDP upper limits, in particular, if social welfare and wellbeing can be maintained or even improved in a shrinking economy as proponents of de-growth suggest (Van Den Bergh, 2017; Van den Bergh, 2011; Van den Bergh and Kallis, 2012). For instance, a steady decrease in population might imply that GDP per capita could be maintained or even increase, at least during a certain amount of time (Cosme et al., 2017; Kerschner, 2010). However, at the current stage, it is difficult to tell if these claims are valid.

What are the possibilities to sustain economic growth beyond the bounds? We cannot escape thermodynamics, but by looking at equation (7), we can still derive several potential strategies. These include (1) to decouple economic activities from energy consumption; (2) to compensate for additional heat released by reducing solar irradiation, and (3) to perform economic activities outside Earth, in space. These 3 strategies which are not mutually exclusive, will be discussed in the following sections.

## 4.1 Decoupling economic activities from energy consumption

A common argument against "limits to growth" is that economic growth can essentially be decoupled from its underlying physical basis and technological progress would enable the decoupling (Mayumi et al., 1998; Stiglitz, 1974). While a detailed discussion of decoupling is beyond the scope of this paper, we will briefly address, how decoupling would impact our results.

According to Ward et al. (2016), decoupling occurs if the rate of economic growth is compensated by a proportional decrease in resource consumption, such that the total resource consumption remains constant. Total decoupling occurs when the rate of decrease in resource consumption is faster than the rate of economic growth, resulting in a decrease in resource consumption. As we are limiting our analysis to energy consumption, we need to compare the rate of decrease in energy consumption with to the rate of economic growth.

Our model does not allow for decoupling as the derivative of the correlation between GDP and energy consumption in equation (1) is positive. This is of course under the condition that this power law based on six decades of data remains valid several centuries into the future.

Although there is no fundamental reason why complete decoupling should be infeasible due to future technology development or growth paths favoring low energy services (Ayres, 2007), more recent empirical evidence does not seem to provide much support for near complete decoupling at the global level (Bithas and Kalimeris, 2013; Ward et al., 2016).

## 4.2 Reducing solar radiation: Geo-engineering

Another approach for circumventing the energy constraints is to reduce solar radiation as a source of global warming. Massive geoengineering approaches might be capable of mitigating global warming by reducing the incoming solar radiation to Earth (Bengtsson, 2006; Boyd, 2008; Keith, 2001; Vaughan and Lenton, 2011). In case heat dissipation becomes the main limiting factor of GDP growth, this could potentially allow for trading low efficient incoming solar radiation (which as we have seen requires a lot of dedicated real estate to convert into large energy generation) for potentially more efficient carbon-neutral energy sources.

The most straightforward way to reduce solar radiation on the Earth's surface is to disperse large amounts of aerosols into the stratosphere, such as sulfate particles (Rasch et al., 2008). No particular technology breakthrough would be required for this approach. However, this type of geoengineering is highly controversial due to difficulties in anticipating its consequences, the high maintenance costs, and the potentially irreversible manipulation of the climate (Barrett, 2008).

In-space geoengineering using space mirrors (Angel, 2006; Mautner, 1993) would decrease on demand the radiation balance by preventing a fraction of incoming sunlight to heat the Earth's atmosphere. A space based approach is certainly much more difficult than the aerosol option, but could potentially provide better flexibility and control on the outcome.

In any case, the overall heat dissipation and therefore thermodynamics of our biosphere would remain unchanged, but more energy could be generated. This would introduce an extra factor in equation (7) that we could rewrite as:

$$\frac{G(t_{lim})}{G(t_0)} = 0.928 \left[ \frac{E_{r\,lim}(1-m)}{E(t_0)} + \frac{(\Delta T_{lim} - \Delta T_{Cmax})}{C}\left(\frac{R_{sol}}{E(t_0)}\right) + m\,\frac{R_{sol}}{E(t_0)} \right]^{1.48} \qquad (8)$$

with m the fraction of solar radiation that can be blocked via geoengineering. The renewable energy limit $E_{r\,lim}$ is reduced by m, under the simplifying assumption that the reduction in solar radiation is evenly distributed across the globe.

It is estimated that a reduction of the incoming sunlight by m = 2% requires a space mirror swarm with a total mass of the order of $10^7$ tons (McInnes, 2010). An endeavor of this type would be particularly relevant in Scenario 3, as it would allow for increasing the upper limit of Earth's economy by a factor

3. While these gains are certainly appreciable, it would provide less than a century of extra meaningful GDP growth.

## 4.3   Performing economic activities in space

A strategy for enabling further GDP growth would be to conduct economic activities increasingly in space. As Martin (1985) has argued, energy supply from space to Earth does not circumvent the heat dissipation limit, as the energy is consumed on Earth. Hence, the energy required to produce products and services has to be consumed in space.

We can imagine two ways of how an economy producing products and services in space might exist:

(a) The economic agents are located on Earth, but part of the value is generated in space. This is already the case, for instance, when people pay for satellite services. The energy required to operate a satellite is consumed in space, but the related service is directly contributing to Earth's GDP. As of today, the energy consumed in space is obviously negligible compared to what is consumed on Earth. However, in the future, as more value is generated in space, this off Earth energy consumption may become substantial.

(b) The economic agents are located off Earth. Therefore, all of the energy required for an economic transaction is consumed in space. Hence, the GDP would be attributed to space. In principle, this would allow for a vast increase in GDP growth, at least from an energy and resource point of view (Hempsell, 1998; Kaneda et al., 1992; Martin, 1985; Vajk, 1976). The challenges to establish a large population of humans off Earth are formidable and beyond the scope of this study.

How large would such a space economy be? Once we reach the limit of Earth's economy, sustaining just 1% GDP growth per year implies that the space economy will become bigger than Earth's economy in just 70 years. After a couple of centuries, the space economy would already dwarf the Earth's economy by an order of magnitude. Kick starting a large space economy will require decades and upfront funding (Hempsell, 1998). To rely on its own momentum, this space economy will need to reach a certain critical size way before we approach Earth's own GDP limit.

Furthermore, if due to excessive global warming from CO2 emissions, our future ability to harvest new energy sources becomes highly constrained, it will drastically limit the maximum size of the terrestrial economy. This smaller economy will have a smaller capacity to invest in research and development, including projects devoted to the development of an in-space economy. This might be an issue, as the required investments for developing a large in-space economy are likely considerable. For example, Kaneda et al. (1992) assume in their scenario that the equivalent of 30% of the world GDP of the year 2000 need to be invested in developing a large-scale space economy. As suggested by current trends in research and development spending,  the fraction of world GDP available for such investments is unlikely to grow beyond a few percent of the world's GDP (World Bank, 2020). It follows that for the scenario from Kaneda et al. (1992), an at least one or two magnitude larger world economy would be required. In this perspective, it is interesting to note that the likelihood to develop a large space economy might be highly impacted by the choices we will make collectively in the coming years regarding greenhouse gas emissions.

## 5   Conclusions

This article showed that the overall size of Earth's global economy is facing an upper limit purely due to energy and thermodynamic factors. Our results indicate that for a 2% annual GDP growth, the upper limit will be reached at best within a few centuries, even in favorable scenarios where new, carbon-neutral energy sources such as fusion power are deployed on a massive scale. We conclude that unless GDP can be largely decoupled from energy consumption, thermodynamics will put a hard cap on the size of Earth's economy. Depending on how fast we decarbonate our economy, this thermodynamic constraint will have major implications on our long-term ability to become a spacefaring civilization.